\documentclass[twocolumn,showpacs,preprintnumbers,amsmath,amssymb,pre]{revtex4}

\usepackage{epsfig}
\usepackage{bm}

\begin{document}


\newcommand{\dd}{\mathrm{d}}
\newcommand{\gcc}{\mbox{g~cm$^{-3}$}}
\newcommand{\kB}{k_\mathrm{B}}
\newcommand{\etal}{{et al.}}
\newcommand{\beq}{\begin{equation}}
\newcommand{\eeq}{\end{equation}}
\newcommand{\bea}{\begin{eqnarray}}
\newcommand{\eea}{\end{eqnarray}}
\newcommand{\req}[1]{Eq.\ (\ref{#1})}
\newcommand{\Znuc}{Z_\mathrm{nuc}}
\newcommand{\Ftot}{F_\mathrm{tot}}


\newcommand{\ApJ}[1]{{Astrophys.\ J.} \textbf{#1}}
\newcommand{\ApJS}[1]{{Astrophys.\ J. Suppl.\ Ser.} \textbf{#1}}
\newcommand{\AandA}[1]{{Astron.\ Astrophys.} \textbf{#1}}
\newcommand{\ARAA}[1]{{Annu.\ Rev.\ Astron.\ Astrophys.} \textbf{#1}}
\newcommand{\PR}[1]{{Phys.\ Rev.} \textbf{#1}}
\newcommand{\PRA}[1]{{Phys.\ Rev. A} \textbf{#1}}
\newcommand{\PRE}[1]{{Phys.\ Rev. E} \textbf{#1}}
\newcommand{\PRL}[1]{{Phys.\ Rev.\ Lett.} \textbf{#1}}


\title{Equation of state for partially ionized carbon 
at high temperatures}

\author{Alexander Y. Potekhin}\email{palex@astro.ioffe.ru}%
\affiliation{Ioffe Physico-Technical Institute,
     194021 St.\ Petersburg, Russia}
     \author{G\'erard Massacrier}\email{Gerard.Massacrier@ens-lyon.fr}
\affiliation{Ecole Normale Sup\'erieure de Lyon,
     CRAL (UMR CNRS No.\ 5574),
     69364 Lyon Cedex 07, France}
\author{Gilles Chabrier}\email{chabrier@ens-lyon.fr}
\affiliation{Ecole Normale Sup\'erieure de Lyon,
     CRAL (UMR CNRS No.\ 5574),
     69364 Lyon Cedex 07, France}

\received{9 December 2004;
revised manuscript received 8 August 2005}

\begin{abstract}
Equation of state for partially ionized carbon at temperatures 
$T\gtrsim 10^5 $~K 
is
calculated 
in a wide range of densities,
using the method of free energy minimization in
the framework of the chemical picture of plasmas. The free energy model
includes the internal partition
functions of bound species.
The latter are calculated by a self-consistent 
treatment of each ionization stage in the plasma environment
taking into account pressure ionization. The
long-range Coulomb interactions between ions and screening of the ions
by free electrons are
included using our previously published analytical model.
\end{abstract}

\pacs{52.25.Kn, 05.70.Ce, 52.27.Gr, 64.30.+t}


\maketitle

\section{Introduction}

The understanding of the physical properties of matter at high
densities and temperatures is important  for the fundamental
physics as well as  for various physical and astrophysical applications.
Since the 1980s the theoretical interest in matter under such unusual
conditions (e.g., Refs.\ \cite{MWYZ88,SBY01,PDWB02,BF04})
has been enhanced by
laboratory developments like
high-power short duration lasers, shock-induced plasmas, inertial confinement
implosions,
or exploding metal wires
(e.g., Refs.\ \cite{GGLetal03,RBCetal02,RWMetal02,ASFetal02,BMTetal02}).
In the astrophysical domain the
  calculation of the equation
of state (EOS) for stellar partial  ionization zones is a
particularly challenging problem. In these zones the electrons and
different ionic species cannot be regarded as simple ideal gases: 
Coulomb interactions, bound-state level shifts, pressure ionization,
 and electron
degeneracy should be taken into account.  In this paper, we
calculate the EOS for carbon at temperatures
$10^5\mbox{~K}\lesssim T \lesssim 10^7 $~K 
in a wide range of
densities $\rho$. Such an EOS is required, e.g., for
modeling inner envelopes of carbon-rich white dwarfs
\citep{FVH76,DAM} 
or outer envelopes of neutron stars.

An EOS calculation in the partial ionization 
regime is not possible without approximations. For astrophysical
simulations, these approximations should not
violate the \textit{thermodynamic consistency}. 
The free energy minimization
method \citep{Harris-ea60,GHR69}
allows one to include the complex
physics in the model and ensures the
consistency.
This
method has the great advantage to 
identify the various contributions to
the free energy, illustrative of various physical effects
(see, e.g.,
Ref.\ \cite{FGV77}, for a
discussion). 

Free-energy models which carefully include the nonideal effects
have been proposed for fluid hydrogen \citep{SC} and helium
\citep{AC94,WC04};
 the EOS tables for these elements,  which cover a
pressure and temperature range appropriate for low-mass stars,
brown dwarfs, and giant planets have been calculated
in \cite{SCVH}.
For heavier elements, a similarly detailed EOS is
lacking. Up to now, the best available thermodynamically
consistent  EOS for carbon covering 
the stellar pressure ionization zones was
the EOS developed by Fontaine, Graboske, and Van Horn in
the
 1970s
\citep{FGV77} (FGV) and amended in 1990s \citep{FGV99}. 
We shall call these two versions FGV77 and
FGV99, respectively. This EOS has been calculated by different
methods in different $\rho-T$ domains. At relatively low densities
(e.g., $\rho<(0.01-1)$ \gcc\ for $10^5 $~K $<T<10^6 $~K), the
ionization equilibrium has been obtained  by the free-energy
minimization technique. At densities above several
grams per cubic centimeter,
 the Thomas-Fermi model has
been employed.  At intermediate densities, in particular 
in the various regimes of pressure ionization, 
the EOS was
interpolated  
between these two regions. Clearly, the
accuracy of the EOS in the interpolation region can be called into
question. Moreover,  the Thomas-Fermi model may be inaccurate
at $\rho\lesssim 10^3$ \gcc, where the pressure is
not sufficiently high to force the complete ionization of carbon,
as we shall see below.

Extension of the free energy minimization technique to 
$\rho\gtrsim0.1$ \gcc\ is complicated because of the growing
importance of nonideal contributions to the free energy and
the onset of pressure
ionization. The latter
is difficult to treat in
the framework of the ``chemical picture'' of plasmas, which assumes that
different ion species can be clearly identified (see, e.g.,
Refs.\ \cite{SCVH,P96,Rogers00}, for discussion). On the other hand,
EOS calculations within the more rigorous ``physical picture,''
quite successful at  relatively low $\rho$ (e.g., \citep{OPAL-EOS}),
become prohibitively complicated  at such high densities.
First principle approaches based on path integral Monte Carlo (PIMC) 
\citep{Bezkr-ea04} or molecular dynamics (MD) calculations 
are computationnaly highly expensive. These methods also suffer from
some difficulties. Indeed, the sign or node problem for the 
PIMC method or the use of effective pair potentials for
MD simulations restrict their applicability
(see however \citep{SBY01}). In any case, a comparison with our
results will be instructive, but, to the best of our knowledge,
no PIMC or MD data for carbon in the temperature-density range of 
interest in this paper has been published yet.

In this paper we present an EOS model which relies on the free
energy  minimization in the framework of the chemical picture
and extends to arbitrarily high densities
across the pressure ionization region \emph{without interpolation}. This
allows us to obtain not only the thermodynamic functions, but also
number fractions for every ionization stage. We
treat the long-range interactions in the system
of charged particles (ions and electrons) using the theory
previously developed for fully ionized plasmas \citep{CP98,PC00}.
The contribution of the internal electronic structure of the ions 
embedded in the dense plasma is calculated using a scheme
\citep{Massacrier} which self-consistently:
(i) builds separate  models for different ionization stages
in the plasma, taking into account the real structure of bound states
(configurations, LS terms);
(ii) uses Boltzmann statistics to sum up the internal partition functions of
these ions;
(iii) takes into account spreading of bound states into energy bands as they
are pressure ionized;
and
(iv) treats quantum mechanically the free electron background around 
each ion thus resolving resonances.
Points (i) and (ii) make our model different from average atom ones.
The closest``ion-in-plasma'' theoretical model is 
that of \citet{PainBlenski},
where ions are treated separately (using 
superconfigurations), but screening is introduced through
a Thomas-Fermi approach for the free electrons. 
The applicability of
our model is tested by numerical calculations of thermodynamic
functions, which we
compare with the FGV models.

In Sec.~\ref{sect-fren} we present the free energy model.
The technique for the calculation of thermodynamic functions 
at equilibrium
is described in Sec.~\ref{sect-TDE}. In Sec.~\ref{sect-res} we
discuss the results of the EOS calculations for carbon plasma,
and in Sec.~\ref{sect-concl} we give conclusions.

\section{Free energy model}
\label{sect-fren}

Consider a plasma consisting of $N_e$ free electrons and
$N_i$ heavy ions
with numbers of bound electrons $\nu$ from 0 to ${\Znuc}$
(where ${\Znuc}$ is the element charge number)
in a volume $V$. 
Let us write
the total Helmholtz free energy as
$
    {\Ftot} = F_{e}
    +F_{i}
     + F_\mathrm{ex},
$
where $F_{i,e}$ 
denote the ideal free energy
of ions and free electrons, respectively, 
and $F_\mathrm{ex}$ is the excess (nonideal)
part, which arises from interactions.
 $F_{i}$ is the free energy of an ideal Boltzmann gas mixture,
which can be written as
$
   F_{i} = N_i \kB T \left[
      \ln(n_i^{\phantom{3}} \lambda_i^3) - 1 \right] -
      S_\mathrm{mix} T,
$
where $\lambda_i = (2\pi\hbar^2/m_i\kB T)^{1/2}$ 
is the thermal wavelength of the ions, $m_i$ is the ion mass,
$
   S_\mathrm{mix} = - N_i \kB \sum_{\nu} x_\nu\ln x_\nu
$
is the entropy of mixing,
and
$x_{\nu} = N_\nu/N_i$ is the number fraction of the ions of the
$\nu$-th type ($\sum_{\nu} x_{\nu} = 1$).
For the electrons at arbitrary degeneracy,
 $F_{e}$ can be expressed through
Fermi-Dirac integrals and approximated by 
analytical formulae \citep{CP98}.
The main complication is the calculation of the nonideal term,
which is quite nontrivial at high densities.
It includes a contribution due to 
the building of localized bound states
of the ions, and the long range 
Coulomb interactions between
these ions and free electrons.
We write
\beq
   F_\mathrm{ex} = F_{ee} + F_{ii} + F_{ie} + F_\mathrm{int}, 
\label{Fex}
\eeq
where the first three terms represent the contributions of
electron-electron, ion-ion, and ion-electron interactions,
respectively, and $F_\mathrm{int}$ is the contribution
due to the internal degrees of freedom of the ions,
that involves sums over bound states. 
Equation (\ref{Fex}) does 
not imply a strict separation of the terms on its right-hand side:
No strict definition of free and bound electrons nor ions exists in a dense plasma.
In general, the terms must be interdependent
and evolve in a correlated way. 
Our approach
to this difficulty consists in 
calculating self-consistent models for the ions
embedded in the plasma, coupling them with a model for the long range interaction,
and minimizing the resulting total free energy $\Ftot$.

\subsection{Free energy of a fully ionized plasma}
\label{sect-fi}

A fully ionized electron-ion plasma which contains only one ion
species is characterized by three parameters: the ion charge $Ze$, 
the electron density parameter $r_s$, and
the ion Coulomb coupling parameter $\Gamma$:
\beq
   r_s = \frac{a_e}{a_0}    ,
\qquad 
   \Gamma = \left(\frac{4\pi\,n_e}{3}\right)^{\!1/3}
        \frac{Z^{5/3} e^2}{\kB T}   ,
\label{Gamma}
\eeq
where 
 $n_e$ is the electron number density and
 $a_0=\hbar^2/m_e e^2$ is the Bohr radius.
The Helmholtz free energy of the fully ionized plasma
is described by analytical fitting formulae \citep{CP98,PC00},
which
are applicable at high
densities ($r_s\lesssim1$, arbitrary $\Gamma$)
or high temperatures (small $\Gamma$). When
neither $r_s$ nor
$\Gamma$ are small, the plasma cannot be considered as fully ionized. 

In a multicomponent fully ionized, dense plasma with different ion
charges $Z_{\nu} e$, the ``linear mixing rule'' has been shown to
be very accurate \citep{HTV77,DWSC96,CA90}:
\beq
   F_\mathrm{ex}^\mathrm{fi}(N_i,V,T,\{x_\nu\}, \{Z_\nu\})
    =  N_i \kB T \sum_{\nu} x_{\nu} 
      f_\nu,
\label{Fex-fi}
\eeq
where 
\beq
 f_\nu = f_\mathrm{ex}^\mathrm{fi}(n_e,T,Z_\nu) =
  F_\mathrm{ex}^\mathrm{fi}\big|_{x_\nu=1}/N_i \kB T
\label{f-ex-def}
\eeq
is obtained from
$
   F_\mathrm{ex}^\mathrm{fi} = F_{ee} + F_{ii} + F_{ie}
\label{Fex-fi3}
$
(the superscript ``fi'' indicates full ionization).
In \req{f-ex-def} $n_e$ takes the value implied by the electroneutrality:
$n_e = n_i \bar{Z}$, where $n_i = N_i/V$
is the ion number density, and $\bar{Z}e=\sum_\nu x_\nu Z_\nu e$
is the mean ion charge.
An effective ion Coulomb parameter for a multicomponent plasma
is obtained by replacing $Z^{5/3}$ with $\sum_\nu x_\nu Z_\nu^{5/3}$
in \req{Gamma} for $\Gamma$.

\subsection{Bound-state contribution to the free energy}

In order to evaluate $F_\mathrm{int}$,
 we calculate the ionic structure in the plasma using the
scheme described in \cite{Massacrier}. It is based on the
ion-sphere approximation, which replaces the actual plasma environment 
for every ion by the statistically averaged boundary conditions for
the electron wave functions within a spherical volume centered at
the ionic nucleus. 
At present we do not include neutral atoms ($\nu={\Znuc}$), 
which is justified at the temperatures and densities
where the ionization degree of the plasma is high.
For each ion
containing $\nu$ bound electrons, 
a radius of the ion sphere $R_\nu$
is determined self-consistently from the requirement
that the sphere is overall electrically neutral. 
The Hamiltonian for the ion $\nu$ is written as
 $H_{\nu} = \sum_{i=1}^\nu h_\nu(\bm{r}_i)+W_\nu$, where
\bea\hspace*{-2em}&&
   h_\nu(\bm{r}) =
   -\frac{\hbar^2}{2}\nabla^2+V_\mathrm{at}^\nu(r)+V_\mathrm{f}^\nu(r),
\label{h-nu}
\\\hspace*{-2em}&&
   W_{\nu} =
   \sum_{i=1}^\nu\left(-\,\frac{{\Znuc}
   e^2}{r_i}- V_\mathrm{at}^\nu(r_i)\right)
     + \sum_{i<j}^\nu\frac{e^2}{|\bm{r}_i-\bm{r}_j|},
\label{W-nu}
\eea
$V_\mathrm{f}^\nu$ is the potential due to the plasma on the ion $\nu$,
that must be determined self-consistently, 
$W_\nu$
is responsible for the $LS$ splitting of spectroscopic terms, and
$V_\mathrm{at}^\nu$ is a scaled Thomas-Fermi potential of the
nucleus and $\nu-1$ bound electrons \citep{EN69}.
Note that $V_\mathrm{at}^\nu$ disappears in $H_\nu$. It is used
to build an effective one-electron Hamiltonian $h_\nu$,
which generates a one-electron wave functions basis.
The coordinate parts $\psi_{nlm}^\nu$ of these functions are obtained
from the Schr\"odinger equation 
\beq
 h_\nu\psi_{nlm}^\nu
  =\epsilon_{\nu nl}\psi_{nlm}^\nu(\bm{r}).
\label{one-el}
\eeq
Then $H_\nu$ is diagonalized in a subspace of Slater determinants
generated by a set of $\psi_{nlm}^\nu$.
The $\nu$-electron energies of the bound states are 
well approximated as
 $
   E_{\nu\alpha} = E_{\nu\alpha}^0 + \sum_{(nl)\in\alpha}
   (\epsilon_{\nu nl}-\epsilon_{\nu nl}^0),
 $
where $E_{\nu\alpha}^0$ and $\epsilon_{\nu nl}^0$ are calculated for the
isolated ion, and  $\alpha=(nl)_1 (nl)_2\ldots(nl)_\nu\,^{2S+1}L$
 defines a particular $LS$ term of a configuration.
The separation of $H_\nu$ into parts (\ref{h-nu}) and (\ref{W-nu})
allows one to capture the plasma effects in one-electron energies
and wave functions through \req{h-nu}, while the $\nu$-electron structure
is retained through the contribution $W_\nu$.
The boundary condition at $R_\nu$ for \req{one-el} does
not noticeably affect $E_{\nu\alpha}$
 except near the densities where the
corresponding term $\alpha$ becomes pressure-ionized. 
The latter case will be addressed below.

The free electron density $n_\mathrm{f}(r)$
and the potential $V_\mathrm{f}^\nu(r) $
are determined self-consistently, using the local density
approximation of the density functional theory.
The one-electron
wave functions $\psi_{\epsilon lm}^\nu(\bm{r})$ 
 of the partial
scattering waves are calculated from the Schr\"odinger
equation 
\bea\hspace*{-1em}&&
   \bigg( -\frac{\hbar^2}{2}\nabla^2 -\frac{{\Znuc}e^2}{r} 
   +V_\mathrm{b}^\nu(r)
\nonumber\\&&\qquad  
   +V_\mathrm{f}^\nu(r)
 +V_\mathrm{xc}^\nu(r) \bigg)
   \psi_{\epsilon lm}^\nu(\bm{r})
    = \epsilon \, \psi_{\epsilon lm}^\nu(\bm{r}).
\label{free-wf}
\eea
Here,
 $V_\mathrm{xc}^\nu$ is the exchange-correlation potential
\citep{PDw84}, $V_\mathrm{b}^\nu$ and $V_\mathrm{f}^\nu$ are
obtained from the Poisson
equation: $\nabla^2 V_\mathrm{b}^\nu = -4\pi n_\mathrm{b}^\nu e^2$,
 $\nabla^2 V_\mathrm{f}^\nu = -4\pi n_\mathrm{f}^\nu e^2$,
and the number densities 
 $n_\mathrm{b}^\nu(r)$ and $n_\mathrm{f}^\nu(r)$ are calculated
as the squared moduli of the wave functions for the bound and free
electrons, respectively, summed with the statistical weights
appropriate for a given $T$. For the bound electrons, 
these weights are proportional to
$w_{\nu\alpha} d_{\nu\alpha} \exp(-E_{\nu\alpha}/\kB T)$, where
$d_{\nu\alpha}=(2S+1)(2L+1)$ is the level degeneracy, and 
$w_{\nu\alpha}$
is an occupation probability defined below.
The density of
states per unit volume for the $l$th partial wave
of the free electrons at a given energy
 $g_{\nu l}(\epsilon)$ is determined with account of the
contribution from resonances (Friedel terms; see
Ref.\ \cite{Massacrier} for details).
The energy distribution of the free electrons
is assumed
 $\propto 
 g_{\nu l}(\epsilon)/(1+\exp[(\epsilon-\mu_e)/\kB T])$. 
The free parameters of the model are $T$ and
the electron chemical potential $\mu_e$.
In thermodynamic equilibrium, $\mu_e$
 is the same for all ions, but as different ionization stages have 
different neutrality sphere radii as well as different numbers of 
neutralizing free electrons inside them, $\mu_e$ can be related 
to the mean free electron density only after the global
free-energy minimization (Sec.~\ref{sect-TDE}) has given 
the relative populations of the ions.

\begin{figure}
\epsfxsize=8cm
\epsffile{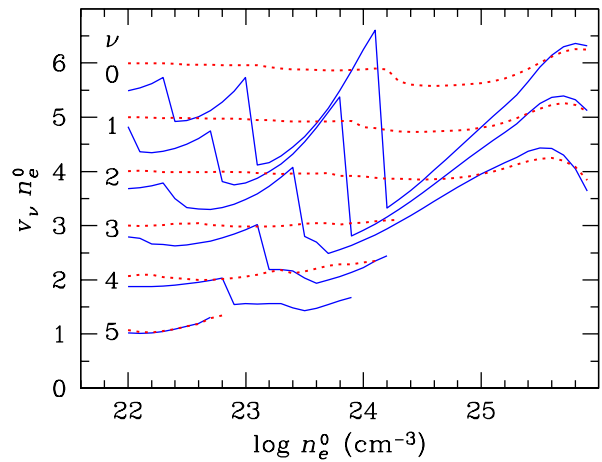}
\caption{Neutrality volumes of carbon ions, $v_\nu$ normalized to
the fiducial number density of free electrons $n_e^0$ (see text),
as functions of $n_e^0$ for $T=2.32\times10^5$~K (solid lines)
and $T=3.71\times10^6$~K (dotted lines) for carbon ions with $\nu$
bound electrons ($\nu=0,1,\ldots,5$). The curves for the three
lowest degrees of ionization ($\nu=3,4,5$) end at the $n_e^0$
values at which there remain no relevant bound states ($\epsilon_-$
become positive
for $2s$ state in $1s^2 2s$ and $1s^2 2s^2$,
 and $2p$ state in $1s^2 2s^2 2p$).
\label{fig-vol}
}
\end{figure}

The neutrality of the ion
 sphere is ensured by the self-consistent determination of $R_\nu$ such
that
\beq
 \frac{\dd}{\dd r} \left[-\frac{\Znuc}{r} + V_\mathrm{b}^\nu(r)
+V_\mathrm{f}^\nu(r)\right]_{r=R_\nu} =0. 
\eeq
Associated with this radius is the \emph{neutrality volume}
$
   v_{\nu} = 4\pi R_\nu^3/3 .
$
In the model of a uniform electron background, that neglects the
interactions of free electrons with ions, one has 
$v_{\nu} = v_{\nu}^0 = ({\Znuc}-\nu)/n_e^0$,
where 
$
  n_e^0(\mu_e,T) = V^{-1}
   \partial F_{e}/\partial\mu_e|_{V,T}
$
is the number density of free electrons
in the uniform gas model.
With allowance for
interactions of the free electrons with the ions and bound
electrons, $v_{\nu}$ deviates from $v_{\nu}^0$,
as illustrated in Fig.~\ref{fig-vol}. The drops of the plotted
curves at certain densities, which are especially sharp at the lower
temperature, are the consequence of pressure ionization of separate
levels: when a $nl$ level of ion $\nu$ 
crosses the continuum limit and appears as
a resonance in the neighboring ionization state $\nu-1$,
the latter ion sphere shrinks to compensate this increase in
the free electron density of states. 

\begin{figure}
\epsfysize=8.5cm
\rotatebox{-90}{\epsffile{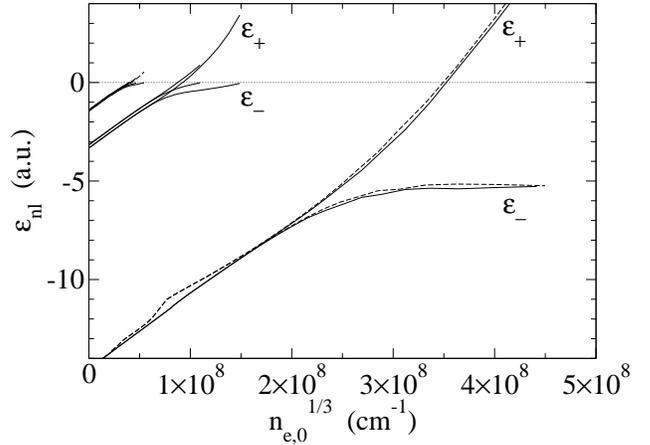}}
\caption{Monoelectronic energies of He-like carbon ($\nu=2$)
as a function of the fiducial electronic density $n_e^0$;
solid line: for $T=3.71\times10^6$~K
 and $1s$, $2s$, $2p$, $3s$, $3p$, $3d$ states
(from bottom to top);
dashed line: for $T=2.32\times10^5$~K and the $1s$ state.
Upper and lower limits of energy bands are marked as
$\epsilon_{+}$ and $\epsilon_{-}$.
\label{fig-energies}
}
\end{figure}

With increasing $\mu_e$ (or $n_e^0$), the radius
 $R_\nu$ decreases,
 the wave functions $\psi_{nlm}^\nu(r)$
 become
 distorted, and
 the energies $\epsilon_{\nu nl}$ 
 spread into a band.
 We estimate a band width by solving \req{one-el} with two
 alternative boundary
 conditions: either $\psi_{nlm}^\nu(R_\nu)=0$, or 
 $\partial \psi_{nlm}^\nu/\partial r=0$ at $r=R_\nu$.
These two
conditions
give two energies which we interpret as the upper
($\epsilon_+$)
and lower ($\epsilon_-$) edges of the band (Fig.~\ref{fig-energies}). 
Eventually $\epsilon_+$ becomes positive.
We interpret the electrons 
with $0 < \epsilon < \epsilon_+$ as quasifree and exclude them from
the internal partition function of the ion. We introduce
an \emph{occupation probability} 
$w_{\nu nl}$, 
equal to the statistical weight of electrons with
$\epsilon<0$ (the significance and thermodynamic meaning
of occupation probabilities in the
chemical picture of plasmas has been discussed, e.g., in
\cite{Fermi24,HM88,P96}).
Assuming for the bands the \citet{Hubbard} density of states
\beq
   g(\epsilon) =
   \frac{2}{\pi\delta^2}\sqrt{\delta^2-(\epsilon-\bar\epsilon)^2},
~~
   \delta=\frac{\epsilon_+-\epsilon_-}{2},
~~
   \bar\epsilon=\frac{\epsilon_+ + \epsilon_-}{2},
\eeq
we obtain, for $\epsilon_- < 0 < \epsilon_+$,
\begin{subequations}  
\bea
   w_{\nu nl} &=& \frac12 - \frac{y}{\pi}\,\sqrt{1-y^2} -
   \frac{1}{\pi}\,\arcsin(y),\qquad
\\&&  
     y = (\epsilon_+ + \epsilon_-)/(\epsilon_+ - \epsilon_-) .
\eea
\end{subequations}  
The occupation probability of a term $\alpha$
is $w_{\nu\alpha} = \prod_{(nl)\in\alpha} w_{\nu nl}$.
For all electron shells $nl$, 
except the $K$ shell, $\epsilon_-$ 
becomes positive at sufficiently high $\mu_e$; in this
case $w_{\nu nl}=0$ and the bound state disappears. 
The lowest curves in Fig.~\ref{fig-vol} end
at the
densities where the bound states  cease to exist in the plasma for
a given $\nu$.

For the $K$ shell of H- and He-like ions, $\epsilon_-$ is negative at any density.
Asymptotically, at large $\mu_e$ (small $R_\nu$) and for a uniform
density $n_e$,
 $\epsilon_- \simeq -0.3\,(\Znuc-\nu)e^2/R_\nu$, while
$\epsilon_+ \simeq 0.5\, m_e^{-1}(\pi\hbar/R_\nu)^2$. Therefore,
at high densities
$w_{\nu nl} \simeq (4\sqrt{6}/\pi^4) [(\Znuc-\nu)R_\nu/5a_0]^{3/2}
\propto n_e^{-1/2}$.

The electrons, that populate the
bands overlapping with the continuum,
are delocalized, and thus add to the free-electron
degeneracy through the Pauli exclusion principle. It means that at a
given $n_i$ the presence of such electrons
increases $\mu_e$. Conversely, at a fixed $\mu_e$, 
the electrons that are
pushed into continuum have a larger neutrality volume, corresponding to
the unscreened shell. Since the
share of these quasifree electrons is $(1-w_{\nu\alpha})$,
an effective ion charge for such
 a 
 partially delocalized state
[to be used in \req{f-ex-def}] is
$
  Z_{\nu} = {\Znuc}-\nu w_{\nu} - \nu^* (1-w_\nu),
$
where
 $w_{\nu}=w_{\nu\alpha}$ for the lowest level $\alpha$ 
of the ion with $\nu$ electrons, 
and
$\nu^* < \nu$ is the number of electrons remaining on the inner
shells ($\nu^*=0$ for $1s$ and $1s^2$).

The contribution of the internal degrees of freedom into the free
energy is calculated as
$
   F_\mathrm{int} = - \sum_{\nu} N_\nu\kB T \ln\mathcal{Z}_\nu,
$
where 
$
   \mathcal{Z}_{\nu} = \sum_{\alpha} 
   w_{\nu\alpha}\,d_{\nu\alpha}\,\exp(-E_{\nu\alpha}/\kB T)
$
is the internal partition function of the
ion in the plasma.

\subsection{Total free energy model}

We evaluate $F_{ee}+F_{ii}+F_{ie}$ in \req{Fex} as described 
in Sec.~\ref{sect-fi}. Albeit this is not strictly
correct for ions with bound states, which are not
pointlike, we need this approximation to make
practical EOS calculations.

The total free energy, normalized to $N_i \kB T$, can be written as
\begin{subequations}
\label{f-tot}
\bea&&\hspace*{-1em}
   f \equiv \frac{{\Ftot}}{N_i \kB T}
    = f^\mathrm{fi} + f_\mathrm{int},
\label{f-tot-main}
\\&&\hspace*{-1em}
\hspace*{-3em}\mbox{where}\quad
   f^\mathrm{fi} = \sum_\nu x_\nu f_\nu + f_{i} + \bar{Z} f_{e},
\label{f-fi}
\\&&\hspace*{-1em}
   f_{i} = \ln(n_i\lambda_i^3) - 1 - s_\mathrm{mix},
\\&&\hspace*{-1em}
   f_{e} = \chi_e - p_e,
\quad 
   \chi_e = \frac{\mu_e}{\kB T},
\quad
   p_e = \frac{P_e}{n_e \kB T},
\\&&\hspace*{-1em}
   f_\mathrm{int} = \frac{F_\mathrm{int}}{N_i\kB T},
\quad
   s_\mathrm{mix} = \frac{S_\mathrm{mix}}{N_i\kB},
\eea
\end{subequations}
and $P_e$ is the free-electron pressure.
All terms of $f^\mathrm{fi}$ can be calculated using the fitting formulae
\citep{CP98,PC00}, and only $f_\mathrm{int}$ should be evaluated
numerically.

\section{Thermodynamic equilibrium}
\label{sect-TDE}

\subsection{Equilibrium conditions}

Thermodynamic equilibrium
at constant $V$ and $T$ realizes at the minimum of the Helmholtz free
energy ${\Ftot}$. Since the total number of the ions
in all ionization states is fixed, this
minimum must be found under the constraint $\sum_\nu N_\nu = N_i$.
The charge neutrality 
condition is satisfied
automatically, because the total number of electrons equals
${\Znuc}$ in each ion cell by construction, however at cost of
the \textit{a priori} unconstrained volume. 
In order to maintain $V=\mbox{constant}$,
one should impose the condition
$\sum_\nu N_\nu v_\nu = V$.

These equilibrium conditions can be written as
\begin{subequations}
\label{eq-min}
\bea&&\hspace*{-2em}
   f = \mathrm{minimum};
\quad
   G_N = G_V = 1;
\quad
   x_\nu\geq0, ~ \forall\nu ;
\\&&\hspace*{-2em}
   G_N = \sum_\nu x_\nu,
\quad
   G_V = n_i \sum_\nu x_\nu v_\nu ,
\eea
\end{subequations}
where
$f=f(\chi_e, T, \{x_\nu\})$ is given by \req{f-tot},
$n_i$ and $T$ are fixed, while $\chi_e$ 
and $x_\nu$ may vary.

\subsection{Finding the equilibrium}
\label{sect-min}

While solving the constrained minimization problem (\ref{eq-min}), we
take into account the condition $G_N=1$ explicitly,
by setting 
$
   x_0=1-\sum_{\nu=1}^{\Znuc-1} x_\nu ,
$
and discard those $\{x_\nu\}$ sets which would result in
the
 negative
right-hand side of this equation.
As mentioned above, we do not consider the neutral atoms
($\nu=\Znuc$).
In order to satisfy the constraint $G_V=1$,
we use the Lagrange multiplier method.
Namely, we minimize an auxiliary function
\beq
   \Phi(n_i,T;\chi_e,\{x_\nu\};\lambda)
   = f - \lambda G_V + \lambda^2 (G_V - 1)^2
\label{Lagrange}
\eeq
with respect to its arguments $\chi_e$ and $x_\nu$ ($1\leq\nu\leq\Znuc-1$)
for different values of the Lagrange multiplier $\lambda$, and find
the $\lambda$ value that gives
$G_V=1$ at the minimum. The last (quadratic) term in \req{Lagrange}
is an empirical regularization
term which accelerates the solution.
The solution provides the equilibrium values of $\Ftot$, $\chi_e$,
 $x_\nu$, and
$n_e=\bar{Z} n_i$.

At each value of $\lambda$, we approach $\min \Phi$ in two stages:
first, a rough position of the minimum is found by the simplex
method,
and then it is refined by the Powell's conjugate-direction procedure
\citep{NRF}.
In order to filter-out
false local minima, the minimization procedure is repeated several
times with different initial sets of parameters, and the
absolute minimum is selected. A search for the root of the equation
$G_V(\lambda)=1$ is performed by bracketing and bisection,
Because of the
complicated dependence of $f$ and $G_V$ on the set of
$x_\nu$, and due to the limited accuracy of 
minimization, $G_V(\lambda)$ may exhibit a numerical
discontinuity, which sometimes disallows the 
bisection to converge,
 so we have
tried several
initial guesses of $\lambda$
  in such cases. 

\subsection{Calculating thermodynamic functions}

Once ${\Ftot}$ is calculated for a range of temperatures and
densities, all thermodynamic functions can be found from its
derivatives. The first derivatives give the pressure
$
 P= - \partial\Ftot/\partial V|_T,
$
entropy $S= - \partial\Ftot/\partial T|_V$,
and internal energy $U=-T^2 \partial/\partial T(\Ftot/T)|_V=F+TS$. 
The second derivatives give,
for example,
 specific heat
$
   C_V = \partial U/\partial T |_V
$
and the pressure exponents (temperature and density logarithmic
derivatives)
$
 \chi_T =
     \partial\ln P / \partial\ln T |_V
$
and
$
 \chi_\rho =
     - \partial\ln P / \partial\ln V |_T.
$
In these derivatives, $N_i$ is kept fixed, but $x_\nu$
depend on $V$ and $T$, following the solution
in Sec.\,\ref{sect-min}.

Although such calculation looks simple, it is technically complicated. 
We achieved the
accuracy of $f$ within 0.003 over the 
$\rho - T$ domain where the electron degeneracy is weak or moderate 
($\chi_e\lesssim10$), 
and to four digits
in the strongly degenerate regime (where $\chi_e\gg1$),
but this is
insufficient for an accurate evaluation of the second 
and mixed derivatives of $f$. The difficulty is partly overcome by
filtering the calculated values. 
We performed
 calculations on a grid
of $(\rho, T)$ points
 and evaluated the derivatives
at each $(\rho,T)$ point from
using the least-squares fit
to the $F$ values at a hundred of neighboring grid points.

This filtering is not sufficient, if the electrons are strongly
degenerate. In this case, the
$T$-derivatives of $\ln\Ftot$, $\ln U$, and $\ln P$ 
are so small that a tiny numerical noise may preclude
their
evaluation. Fortunately, in this
regime these derivatives are mainly determined by $f^\mathrm{fi}$.
           We use
the following
modification of \req{f-tot-main}:
\beq
   f = f^\mathrm{fi} + s_\mathrm{mix} + f' ,\qquad
   f' = \sum_\nu x_\nu \ln(x_\nu/\mathcal{Z}_\nu).
\eeq
The values of $f'$, $x_\nu$, and $\bar{Z}$, and their $\rho$- and 
$T$-derivatives are calculated numerically, as described above,
whereas
$f^\mathrm{fi}$ and its derivatives
are obtained from
the analytical fits \citep{CP98,PC00}.

\begin{figure}
\epsfxsize=85mm
\epsffile{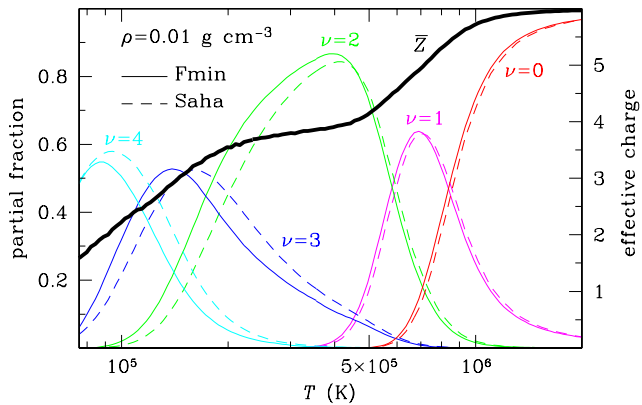}
\caption{Number fractions of different 
carbon ions in the plasma (left vertical axis) 
versus temperature
at \ 
$\rho=10^{-2}$ \gcc.
Solid lines: accurate results; dashed lines: Saha approximation
with current partition functions.
Numbers of bound electrons $\nu$ are marked near the curves.
The thick
solid curve shows the mean effective charge $\bar{Z}$ (right axis).}
\label{fig-z_saha_a}
\end{figure}

\begin{figure}
\epsfxsize=85mm
\epsffile{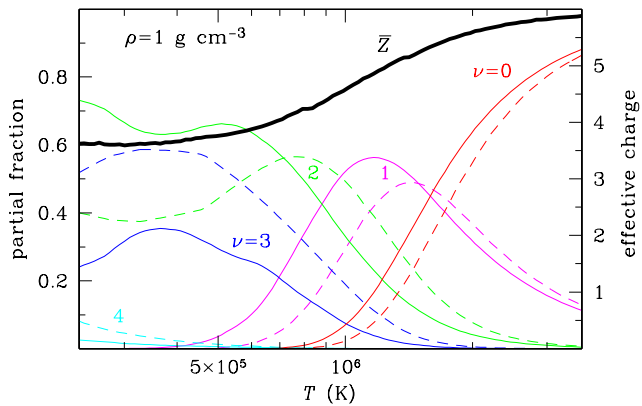}
\caption{The same as in Fig.~\ref{fig-z_saha_a}, but at
$\rho=1$ \gcc.
}
\label{fig-z_saha1}
\end{figure}

\begin{figure}
\epsfxsize=85mm
\epsffile{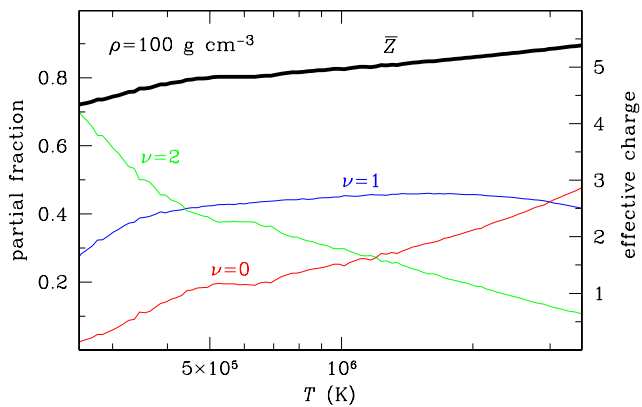}
\caption{Number fractions (left axis) 
and
the mean effective charge $\bar{Z}$ (right axis)
versus temperature
at 
$\rho=100$ \gcc.
}
\label{fig-zt}
\end{figure}

The calculated functions $C_V$, $\chi_T$, and $\chi_\rho$
still exhibit a considerable
numerical noise.
To suppress it, we again employ the least-squares filtering.
Improved values of pressure, consistent with the filtered 
$\chi_T$, are obtained by numerical integration
of the equation $\ln P=\int\chi_T\,\dd t$, starting from the lowest
isotherm.

The thermodynamic stability ($C_V>0$, $P>0$)
and normality ($\chi_T > 0$) require that 
$S/N_i\kB$
monotonically increases with decreasing $\rho$
or increasing $T$.
To maintain these properties, we calculate $S$ by
integration of the equations
$\partial S/\partial V |_T = \chi_T P/T$ and
    $\partial S/\partial T |_V = C_V/T$,
 starting from the
highest $\rho$ and lowest $T$.

\section{Results for carbon}
\label{sect-res}

\begin{figure}
\epsfxsize=85mm
\epsffile{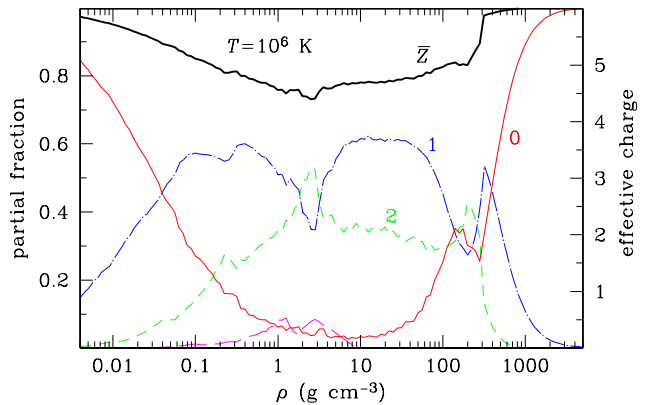}
\caption{
Same
as in Fig.~\protect\ref{fig-zt}, but  
versus density
at constant $T=10^6$~K.
}
\label{fig-zr}
\end{figure}

We have calculated the EOS for carbon at 
$2.34\times10^5\mbox{\,K}\leq T \leq 3.63\times10^6$~K
and $10^{-4}~\gcc \leq \rho \leq 10^4$ \gcc; 
at still higher $\rho$
the carbon plasma is fully ionized. For lower temperatures,
$7.5\times10^4\mbox{\,K}\leq T \leq 2.34\times10^5\mbox{\,K}$,
we have calculated the EOS at 
$10^{-4}~\gcc \leq \rho \leq 0.04$ \gcc.

Figures \ref{fig-z_saha_a}--\ref{fig-zt} show
the $T$-dependences of ion number
fractions $x_\nu$ and the mean effective charge $\bar{Z}$. At the
lower densities, the electrons are nondegenerate. 
In this case, the mean ionization
degree and $\bar{Z}$ depend sensitively on temperature. At
the high density (Fig.~\ref{fig-zt}), the electron degeneracy is
significant, and the number of free electrons
is mainly controlled by pressure, rather than temperature, so that
$\bar{Z}$ varies weakly.
  However, the state of the
 \emph{bound} electrons still depends appreciably on $T$: most of
 them are in the $1s$ state at the higher $T$ and in the $1s^2$
 state at the lower $T$.

The nonideality effects are less important 
at lower density. Therefore, the abundance of individual ion species
at low densities can be evaluated from the Saha equation
\beq
   \frac{x_{\nu+1}}{x_{\nu}} =
    \frac{\mathcal{Z}_{\nu+1}}{\mathcal{Z}_{\nu}} \,
    \frac{n_e^{\phantom{3}}}{2}
    \left(\frac{2\pi\hbar^2}{m_e\kB T}\right)^{\!3/2},
\label{Saha}
\eeq
as illustrated by Fig.~\ref{fig-z_saha_a} for $\rho=0.01$ \gcc. 
Note that 
the shifts of bound-state levels in the plasma environment
are included in $\mathcal{Z}_{\nu}$.
At $\rho\ll 0.01$ \gcc, this approximation and our calculations
give identical results (this is one of the checks of our 
calculations), but
at $\rho\gtrsim0.01$ \gcc, \req{Saha}
becomes progressively inaccurate (Fig.~\ref{fig-z_saha1}). 
The differences between Saha and our models
in Figs.~\ref{fig-z_saha_a} and \ref{fig-z_saha1} are due to
the configurational effects (i.e., the deviations
of the neutrality volumes from their ideal values;
see Fig.~\ref{fig-vol}) and the Coulomb plasma nonideality
(Sec.~\ref{sect-fi}).

The $\rho$ dependences of the ionization states at $T=10^6$~K
are shown in
Fig.~\ref{fig-zr}.
They exhibit
pronounced maxima and minima due to the
pressure ionization of particular bound states 
in particular ionization stages.
These features are related to swelling and
shrinking of 
the individual neutrality volumes relative to their
rigid-background
values (Fig.~\ref{fig-vol}) and the corresponding changes in the
internal partition functions.

\begin{figure}
\epsfxsize=78mm
\epsffile{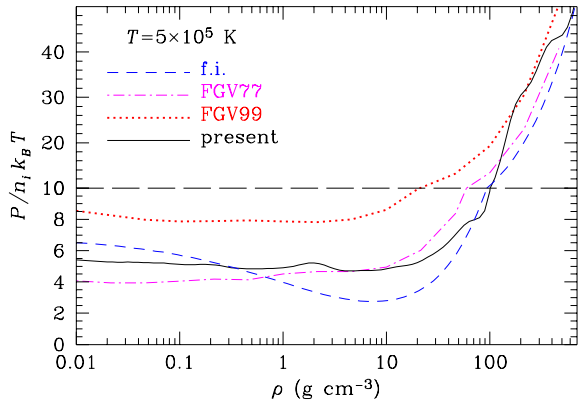}
\caption{Isotherms of normalized pressure $P/n_i\kB T$ for 
$T  = 5\times10^5$~K.
The present data (solid lines) are compared with the FGV77
(dot-dashed lines), FGV99 (dotted lines),
and fully ionized plasma (dashed lines) models.
Note the different scale  in the figure above and below
the horizontal long-dash line.
}
\label{fig-p}
\end{figure}

\begin{figure}
\epsfxsize=77mm
\epsffile{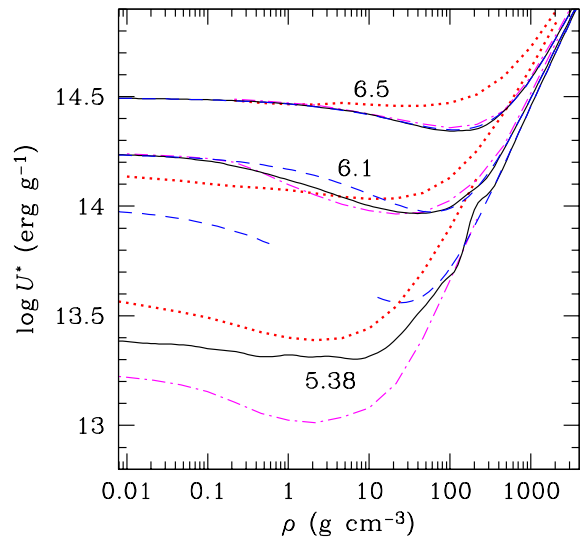}
\caption{Isotherms of internal energy for 
$T = 2.4\times10^5$~K, $1.26\times10^6$~K, and $3.16\times10^6$~K
(the curves are marked by $\log\,T$ values).
The present data (solid lines) are compared with the FGV77
(dot-dashed lines), FGV99 (dotted lines),
and fully ionized plasma (dashed lines) models.
}
\label{fig-u}
\end{figure}

\begin{figure}
\epsfxsize=80mm
\epsffile{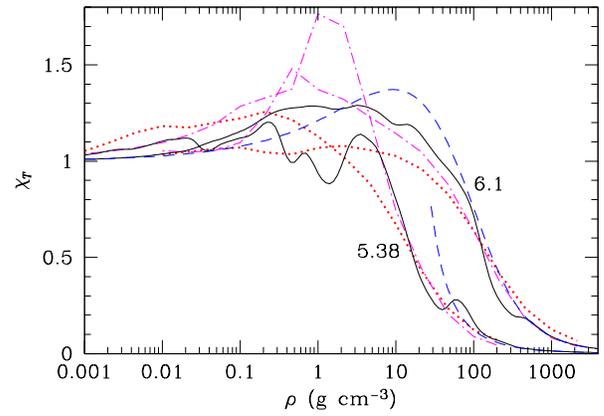}
\caption{Temperature exponent $\chi_T$ for isotherms
$T=2.4\times10^5$~K and $1.26\times10^6$~K, compared with the FGV77
(dot-dashed lines), FGV99 (dotted lines),
and fully ionized plasma (dashed lines) models
(the curves marked with $\log\,T$ values).
}
\label{fig-chit}
\end{figure}

Figure \ref{fig-p} presents normalized
pressure as a function of density at $T=5\times10^5$~K.
The vertical scale is smaller for the upper part of the figure,
to take account of the rapidly growing pressure of degenerate electrons.
The difference between our results and the FGV99 tables
is in general of the same magnitude as the
difference between FGV77 and FGV99. However, 
our isotherms exhibit more features. 
The slope of each isotherm varies near
the densities where the ion composition of the plasma rapidly
changes. These variations could not be revealed by the Thomas-Fermi
model, but are easily grasped within the free-energy minimization
technique. Related variations are seen
in Fig.~\ref{fig-u}, which shows
 isotherms of the internal energy per unit
mass, $U^*=(U+U_0)/N_i m_i$  measured from the energy
level $-U_0$ of a
nonionized ground-state carbon, which corresponds to
a shift equal to  
$8.28\times10^{13}$ erg g$^{-1}$
with respect to the electron continuum level.
The gap in the cold isotherm of the fully ionized plasma model
(dashed line) corresponds to the region
of instability of this model.
Variations of the EOS due to the changing plasma composition
with increasing density are also seen for the temperature derivative
$\chi_T$, shown in Fig.~\ref{fig-chit}.

\begin{figure}
\epsfxsize=80mm
\epsffile{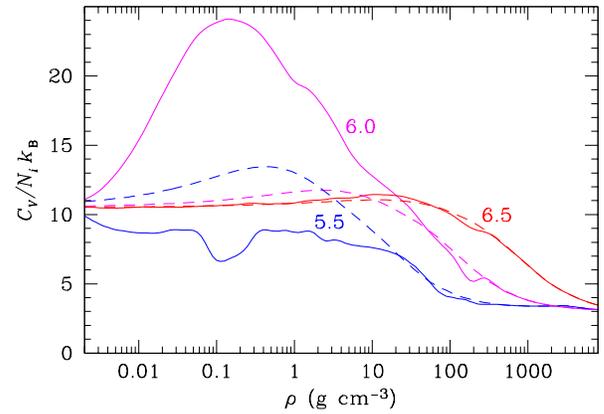}
\caption{Normalized specific heat at constant volume 
$c_V=C_V/N_i\kB$ for
$T = 3.16\times10^5$~K, $10^6$~K, and $3.16\times10^6$~K
(the curves marked with $\log\,T$ values).
Present data (solid lines) are compared with the fully ionized
plasma model (dashed lines).
}
\label{fig-cv}
\end{figure}

Figure \ref{fig-cv} shows the normalized specific heat 
$c_V\equiv C_V/N_i\kB$ as a function of $\rho$ for three values of $T$.
At low $\rho$, where the electrons are nondegenerate, the ideal-gas
value of $c_V$ is  $c_V^\mathrm{id}=1.5(\bar{Z}+1)$.
Actually $c_V$ approaches 
$c_V^\mathrm{id}$ at
$T\gtrsim3\times10^6$~K, where the ionization is almost complete but
in general, these two quantities
can differ considerably, in particular because the heat
absorbed by a partially ionized medium is spent not only on the
increase of the kinetic energy of the ions and electrons, but also
on thermal ionization.
At very high densities, the specific heat is mostly determined by
the ions. 
For a classical ion solid $c_V=3$; for a strongly coupled
ion liquid $c_V\approx3.4$ near the melting $\Gamma$ (Eq.~(17) of
Ref.\ \cite{PC00}). The corrections due to the ion-electron
interactions and quantizing ion motions are within 10\% in 
the high-density part of Fig.~\ref{fig-cv}. 
This explains the limit
 $c_V\approx3$ seen at high $\rho$.

  At $\rho\gtrsim1000~\gcc$, our model reproduces the EOS for a fully
 ionized nonideal plasma \citep{PC00}.

\vspace*{-2ex}
\section{Conclusions}
\label{sect-concl}
\vspace*{-2ex}

We have developed a model for calculation of the EOS for 
dense, partially ionized plasmas, based on the free energy
minimization method and suitable for pressure ionization zones. The
free energy model is
constructed in the framework of the chemical picture of plasmas and
includes detailed self-consistent accounting of quantum states of
partially ionized atoms in the plasma environment. Occupation
probability formalism based on the energy bands
is used to treat pressure ionization. 

The developed technique is applied to the carbon plasma at
temperatures $T\gtrsim10^5$~K, which is
relevant for inner envelopes of the carbon-rich
white dwarfs or outer accreted envelopes of the neutron stars.
For general astrophysical applications of various type of stars,
it is desirable to extend the calculated EOS to 
other chemical elements, first of all to oxygen.
We are planning to perform such calculations in near future.

\vspace*{-2ex}
\begin{acknowledgments}
\vspace*{-2ex}
The authors are grateful to G.~Fontaine for providing updated tables
of the FGV EOS.
A.Y.P.\ acknowledges the hospitality
of the theoretical astrophysics group at the Ecole Normale
Sup\'erieure de Lyon. 
The work of A.Y.P.\ and G.C.\ was partially supported
by the CNRS French-Russian Grant No.\ PICS 3202.
The work of A.Y.P.\ was also supported in part 
the RLSS Grant No.\ 1115.2003.2 
and the RFBR Grant Nos.\ 05-02-16245, 03-07-90200, and 05-02-22003. 
\end{acknowledgments}



\begin{thebibliography}{99}

\bibitem[More \etal(1988)More, Warren, Young, and Zimmerman]{MWYZ88}
R.~M. More, K.~H. Warren, D.~A. Young, and G.~B. Zimmerman,
Phys. Fluids \textbf{31}, 3059 (1988).

\bibitem[Surh \etal(2001)Surh, Barbee, and Yang]{SBY01}
M.~P. Surh, T.~W. Barbee III, and L.~H. Yang,
\PRL{86}, 5958 (2001).

\bibitem[Perrot \etal(2002)Perrot, Dharma-wardana and Benage]{PDWB02}
F. Perrot, M.~W.~C. Dharma-wardana, and J. Benage,
\PRE{65}, 046414 (2002).

\bibitem[Blancard and Faussurier(2004)]{BF04}
C. Blancard and G. Faussurier,
\PRE{69}, 016409 (2004).

\bibitem[Glenzer \etal(2003)]{GGLetal03}
S.~H. Glenzer, G. Gregori, R.~W. Lee, F.~J. Rogers, S.~W. Pollaine, and O.~L. Landen,
{Phys.\ Rev.\ Lett.} \textbf{90}, 175002 (2003).

\bibitem[Renaudin \etal(2003)]{RBCetal02}
P. Renaudin, C. Blancard, J. Clerouin, G. Faussurier, P. Noiret, and V. Recoules,
{Phys.\ Rev.\ Lett.} \textbf{91}, 075002 (2003).

\bibitem[Riley \etal(2002)]{RWMetal02}
D. Riley, I. Weaver, D. McSherry, M. Dunne, D. Neely, M. Notley, and E. Nardi,
\PRE{66}, 046408 (2002).

\bibitem[Audebert \etal(2002)]{ASFetal02}
P. Audebert, R. Shepherd, K. B. Fournier, O. Peyrusse, D. Price, R. Lee,
P. Springer, J.-C. Gauthier, and L. Klein,
{Phys.\ Rev.\ Lett.} \textbf{89}, 265001 (2002).

\bibitem[Batani \etal(2002)]{BMTetal02}
D. Batani, A. Morelli, M. Tomasini, A. Benuzzi-Mounaix, F. Philippe, M. Koenig,
B. Marchet, I. Masclet, M. Rabec, C. Reverdin, R. Cauble, P. Celliers, G. Collins,
L. DaSilva, T. Hall, M. Moret, B. Sacchi, P. Baclet, and B. Cathala,
{Phys.\ Rev.\ Lett.} \textbf{88}, 235502 (2002).

\bibitem[Fontaine and Van Horn(1976)]{FVH76}
G. Fontaine and H.~M. Van Horn,
\ApJS{35}, 293 (1976).

\bibitem[D'Antona and Mazzitelli(1990)]{DAM}
 F. D'Antona and I. Mazzitelli,
\ARAA{28}, 139 (1990).

\bibitem[Harris \etal(1960)Harris, Roberts, and Trulio]{Harris-ea60}
 G.~M. Harris, J.~E. Roberts and J.~G. Trulio,
\PR{119}, 1832 (1960).

\bibitem[Graboske \etal(1969)Graboske, Harwood, and Rogers]{GHR69}
 H.~C. Graboske, Jr., D.~J. Harwood, and F.~J. Rogers,
\PR{186}, 210 (1969).

\bibitem[Fontaine \etal(1977)Fontaine, Graboske, and Van Horn]{FGV77}
G., Fontaine,  H.~C. Graboske, Jr., and H.~M. Van Horn,
\ApJS{35}, 293 (1977).

\bibitem{SC}
D. Saumon and G. Chabrier,
\PRA{44}, 5122 (1991);
\textbf{46}, 2084 (1992).

\bibitem[Aparicio and Chabrier(1994)]{AC94}
 J.~M. Aparicio and G. Chabrier,
\PRE{50}, 4948 (1994).

\bibitem[Winisdoerffer and Chabrier(2004)]{WC04} 
 C. Winisdoerffer and G. Chabrier,
\PRE{71}, 026402 (2005).

\bibitem[Saumon \etal(1995)Saumon, Chabrier, and Van Horn]{SCVH}
D. Saumon, G. Chabrier, and H.~M. Van Horn,
\ApJS{99}, 713 (1995).

\bibitem[Fontaine(1999)]{FGV99}
G. Fontaine (private communication, 1999).

\bibitem[Potekhin(1996)]{P96}
 A.~Y. Potekhin,
Phys.\ Plasmas \textbf{3}, 4156 (1996).

\bibitem[Rogers(2000)]{Rogers00}
F.~J. Rogers,
Phys.\ Plasmas, \textbf{7}, 51 (2000).

\bibitem[Rogers \etal(1996)Rogers, Swenson, and Iglesias]{OPAL-EOS}
F.J. Rogers, F.J. Swenson, and C.A. Iglesias, 
\ApJ{456}, 902 (1996).

\bibitem[Bezkrovniy \etal(2004)]{Bezkr-ea04}
V. Bezkrovniy, V. S. Filinov, D. Kremp, M. Bonitz, M. Schlanges, W. D. Kraeft,
P. R. Levashov, and V. E. Fortov,
\PRE{70}, 057401 (2004).

\bibitem[Pain and Blenski(2003)]{PainBlenski}
 J.-C. Pain and T. Blenski,
  J.~Quant.\ Spectrosc.\ Radiat.\ Trans, 
  \textbf{81},
355 
 (2003).


\bibitem[Chabrier and Potekhin(1998)]{CP98}
G. Chabrier and A.~Y. Potekhin,
\PRE{58}, 4941 (1998).

\bibitem[Potekhin and Chabrier(2000)]{PC00}
A.~Y. Potekhin and G. Chabrier,
\PRE{62}, 8554 (2000).

\bibitem[Massacrier(1994)]{Massacrier}
G. Massacrier,
J.\ Quant.\ Spectrosc.\ Radiat.\ Transfer \textbf{51}, 221 (1994).

\bibitem[Hansen \etal(1977)Hansen, Torrie, and Vieillefosse]{HTV77}
 J.~P. Hansen, G.~M. Torrie, and P. Vieillefosse,
\PRA{16}, 2153 (1977).

\bibitem[DeWitt \etal(1996)DeWitt, Slattery, and Chabrier]{DWSC96}
H. DeWitt, W. Slattery, and G. Chabrier,
Physica B \textbf{228}, 158 (1996).

\bibitem[Chabrier and Ashcroft(1990)]{CA90}
G. Chabrier and N.~W. Ashcroft,
\PRA{42}, 2284 (1990).

\bibitem[Eissner and Nussbaumer(1969)]{EN69}
W. Eissner and H. Nussbaumer,
J.\ Phys. B \textbf{2}, 1028 (1969).

\bibitem[Perrot and Dharma-wardana(1984)]{PDw84}
F. Perrot and M.~W.~C. Dharma-wardana,
\PRA{30}, 2619 (1984).

\bibitem[Fermi(1924)]{Fermi24}
E. Fermi,
Z.\ Phys. \textbf{26}, 54 (1924).

\bibitem[Hummer and Mihalas(1988)]{HM88}
 D.~G.Hummer and D. Mihalas,
\ApJ{331}, 794 (1988).

\bibitem[Hubbard(1964)]{Hubbard}
J. Hubbard,
Proc.\ R.\ Soc. London, Ser.\ A \textbf{281}, 401 (1964).

\bibitem[Press \etal(1992)]{NRF}
 W.~H. Press, S.~A. Teukolsky, W.~T. Vetterling, and B.~P. Flannery, 
\textit{Numerical Recipes in Fortran
}, 2nd ed.
(Cambridge University Press, Cambridge, UK, 1992).

\end{thebibliography}
\end{document}